\documentclass[12pt]{article}
\input{epsf}

\begin{document}

\begin{titlepage}

\hfill\vbox{
\hbox{CERN-TH/99-153}
\hbox{ULB-TH/99-10}
\hbox{hep-th/9906147}
\hbox{June 1999} 
}\par

\vspace{.5in}

\begin{center}

{\Large \bf
Effective field theory approach to ${\cal N}=4$
supersymmetric Yang-Mills at finite temperature
}\vspace{.5in}

Agustin Nieto~\footnote{\tt Agustin.Nieto@cern.ch}
\\
{\it CERN -- Theory Division, CH-1211 Geneva 23,
        Switzerland}
\\ 
and 
\\
Michel H.G. Tytgat~\footnote{\tt mtytgat@ulb.ac.be}
\\
{\it Service de Physique Th\'eorique, CP225, Universit\'e Libre
  de Bruxelles, Bld du Triomphe, 1050 Brussels, Belgium}

\end{center}
\vspace{.5in}
\vfill
\begin{abstract}
  We study the perturbation expansion of the free energy of
  ${\cal N}=4$ supersymmetric $SU(N)$ Yang-Mills at finite
  temperature in powers of 't Hooft's coupling $g^2 N$ in
  the large $N$ limit. Infrared divergences are controlled
  by constructing a hierarchy of two 3 dimensional effective
  field theories. This procedure is applied to the
  calculation of the free energy to order $(g^2 N)^{3/2}$,
  but it can be extended to higher order corrections.
\end{abstract}
\vfill

\end{titlepage}

In recent years some light has been shed on understanding
the relation between string theory and gauge
theories~\cite{Klebanov:1999ku}. A particularly interesting
system is $N$ coincident, parallel D3-branes.  This system
realizes ${\cal N}=4$ supersymmetric $U(N)$ Yang-Mills in
its world volume (4 dimensions). In the large $N$ limit, the
3-brane system becomes a black brane whose
Bekenstein-Hawking entropy can be obtained by considering
Ramond-Ramond charged 3-brane classical
solutions~\cite{Gubser:1996de}. It is therefore interesting
to compare the 3-brane thermodynamics with that of ${\cal
  N}=4$ supersymmetric $U(N)$ Yang-Mills. The free energy
density of ${\cal N}=4$ supersymmetric $U(N)$ Yang-Mills in
the ideal gas approximation is ${\cal F}_{\rm ideal}=-\pi^2
N^2 T^4 /6$. On the other hand, the free energy of a black
3-brane was found to be~\cite{Gubser:1996de} ${\cal F}_{\rm
  BH}=(3/4)\times{\cal F}_{\rm ideal}$.  Maldacena's
conjecture~\cite{Maldacena:1997re} helps to understand the
relative factor $3/4$ between ${\cal F}_{\rm ideal}$ and
${\cal F}_{\rm BH}$.  The conjecture relates type IIB
superstrings on ${\rm AdS}_5\times{\rm S}^5$ to ${\cal N}=4$
supersymmetric $SU(N)$ Yang-Mills in 4 dimensions and allows
to compute the next-to-leading order correction to ${\cal
  F}_{\rm BH}$ in the strong coupling
limit~\cite{Gubser:1998nz},
\begin{equation}
  {{\cal F}_{\rm BH}\over{\cal F}_{\rm ideal}} =  {3\over 4}
     + {45\zeta(3)\over 64\sqrt{2}}{1\over\lambda^{3/2}} \,,
\label{FBH}
\end{equation}
where $\lambda = g^2 N$ is the 't~Hooft coupling.
$f(\lambda)\equiv{\cal F}/{\cal F}_{\rm ideal}$ should be
interpreted as a function whose strong coupling limit is
$f(\infty)=3/4$, while the weak limit is $f(0)=1$. It has
been suggested that $f(\lambda)$ is a monotonic function that
interpolates between the strong and the weak coupling
limits~\cite{Klebanov:1999ku,Gubser:1998nz}.

We will explore the weak coupling limit of ${\cal N}=4$
supersymmetric $SU(N)$ Yang-Mills (SYM for short) in 4
dimensions at finite temperature (3 space dimensions). In
the present note, we will carry out the calculation of
$f(\lambda)$ up to order $\lambda^{3/2}$ by using the
effective field theory approach at finite
temperature~\cite{Farakos:1994kx,Braaten:1995cm}.  Related
work can be found in~\cite{Fotopoulos:1998es}
and~\cite{Vazquez-Mozo:1999ic,Kim:1999sg}, where the terms
of order $\lambda$ and $\lambda^{3/2}$ have also been
computed, respectively, but using approaches different from
ours.

\bigskip

Feynman rules at finite $T$ are the same as those at $T=0$
except that loops involve infinite sums over Fourier
modes~\cite{lebellac}.  The thermodynamic properties of
${\cal N}=4$ supersymmetric $SU(N)$ Yang-Mills in 4
dimensions, in thermal equilibrium are described by the free
energy density ${\cal F} = -T\log{\cal Z}/V$, where ${\cal
  Z}$ is the partition function and $V$ is the 3d volume of
the system, which is taken to infinity in the thermodynamic
limit. The partition function can be written as a path
integral over the fields of the Lagrangian ${\cal L}$ that
describes SYM.  Fields are in the adjoint representation of
the gauge group, the gauge coupling is $g$. An expansion for
$-T\log{\cal Z}/V$ in powers of the coupling constant $g$
can be obtained by summing up the vacuum Feynman diagrams
that contribute at a given order in $g$.

The infrared (IR) behavior of field theories at finite $T$
however prevents us from computing the free energy density
in the form just described. There are two sources of IR
divergences. First, some fields that are massless at $T=0$,
become massive at $T\neq 0$ with a mass of order $gT$.
Diagrams with self-energy insertions have IR divergences
that do not cancel order by order in the coupling constant.
A second source of IR divergences, which is specific to
non-abelian gauge theories, has to do with the gauge field
self-interaction terms.

The effective field theory approach allows to remove
systematically the IR divergences associated to thermal mass
insertions. Also, IR divergences related with the gauge
field self-coupling terms are isolated and interpreted as a
source of non-perturbative effects~\cite{Braaten:1996jr}.
In the effective field theory approach for gauge theories, a
hierarchy of two effective field theories in 3 dimensions
which only contain static modes are constructed. Since, in
addition, these effective field theories do not contain
fermionic fields, they are very convenient for
non-perturbative studies on the
lattice~\cite{Farakos:1994kx}.\footnote{Several theories at
finite temperature have been studied perturbatively and
non-perturbatively using the effective field theory
approach. The $\phi^4$-theory is studied
in~\cite{Braaten:1995cm}, QCD
in~\cite{Braaten:1995na,Braaten:1996jr}, QED
in~\cite{Andersen:1996ej}, and scalar QED in~\cite{SQED}.
The electroweak phase transition has been studied in the
Standard Model~\cite{Farakos:1994kx,EWSM} and in some of
the Minimal Supersymmetric extensions of the Standard
Model~\cite{MSSM}.}

\bigskip

If non-static modes are integrated out in the partition
function ${\cal Z}$, we can write
\begin{equation}
  {\cal Z} = 
  \int {\cal D}A_0({\bf x}) {\cal D}A_i({\bf x}) 
       {\cal D}\phi_I({\bf x}) \;
  e^{- \int d^3 x {\cal L}_{\rm eff}} \, ,
\label{eleZ}
\end{equation}
where ${\cal L}_{\rm eff}$ is an effective Lagrangian
compatible with the internal symmetries of ${\cal L}$.
${\cal L}_{\rm eff}$ describes a 3 dimensional effective
field theory~\cite{drTemp}. ${\cal L}_{\rm eff}$ is made out
of an electrostatic field $A_0({\rm x})$, a magnetostatic
gauge field $A_i({\rm x})$, and 6 scalars $\phi_I({\rm x})$,
where $I=1,\ldots,6$. These fields can be identified, up to
field redefinitions, with the static modes of their 4
dimensional counterparts. Note that fermion fields have been
integrated out completely because they do not have a static
mode.  We can write ${\cal L}_{\rm eff} = f_E + {\cal
  L}_{\rm ESYM}$, where $f_E$ is a constant that cannot be
ignored when computing the free energy. ${\cal L}_{\rm
  ESYM}$ contains the remaining terms and has the form
\begin{eqnarray}
  {\cal L}_{\rm ESYM} &=&
    {1 \over 4} \, G_{ij}^a G_{ij}^a
    \;+\; {1 \over 2} \, (D_i A_0)^a (D_i A_0)^a
    \;+\; {1 \over 2} \, (D_i \phi_I)^a (D_i \phi_I)^a  
\nonumber \\ &&
    \;+\; {1 \over 2} m_E^2 \, A_0^a A_0^a
    \;+\; {1 \over 2} m_S^2 \, \phi_I^a \phi_I^a
    \;+\; \delta {\cal L}_{\rm ESYM},
\label{LESYM}  
\end{eqnarray}
where $G_{ij}^a = \partial_i A_j^a - \partial_j A_i^a + g_E
f^{abc} A_i^b A_j^c$ is the magnetostatic field strength
with gauge coupling constant $g_E$. The term $\delta {\cal
  L}_{\rm ESYM}$ in~(\ref{LESYM}) represents an infinite
number of terms constructed out of $A_0({\bf x})$, $A_i({\bf
  x})$, and $\phi_I({\bf x})$ that contribute only at order
higher than $\lambda^{3/2}$.  $\delta {\cal L}_{\rm ESYM}$
contains non-renormalizable interaction terms. We call the
effective theory constructed by integrating out non-static
modes {\em Electrostatic\/} SYM (ESYM). The free energy can
be written as ${\cal F} = T \left( f_E - \log{{\cal Z}_{\rm
        ESYM}}/ V \right)$, where
\begin{equation}
  {\cal Z}_{\rm ESYM} = 
  \int {\cal D}A_0({\bf x}) {\cal D}A_i({\bf x}) 
       {\cal D}\phi_I({\bf x})  \;
  e^{- \int d^3 x\; {\cal L}_{\rm ESYM}} \, .
\label{ZESYM}
\end{equation}

We can go a step further by integrating out $A_0({\bf x})$,
and $\phi_I({\bf x})$ in~(\ref{ZESYM}). We obtain
\begin{equation}
  {\cal Z}_{\rm ESYM} = 
  \int {\cal D}A_i({\bf x}) \;
  e^{- \int d^3 x \; \widetilde{{\cal L}_{\rm eff}}} \, ,
\label{magZ}
\end{equation}
where $\widetilde{{\cal L}_{\rm eff}}$ is an effective
Lagrangian compatible with the internal symmetries of the
theory which is made out of a magnetostatic gauge field
$A_i({\bf x})$. We can write $\widetilde{{\cal L}_{\rm eff}}
= f_M + {\cal L}_{\rm MSYM}$, where $f_M$ is a constant.
${\cal L}_{\rm MSYM}$ is of the form
\begin{equation}
  {\cal L}_{\rm MSYM} =
    {1 \over 4} \, H_{ij}^a H_{ij}^a 
    \;+\; \delta {\cal L}_{\rm MSYM},
\label{LMSYM}  
\end{equation}
where $H_{ij} = \partial_i A_j^a - \partial_j A_i^a + g_M
f^{abc} A_i^b A_j^c$ is the magnetostatic field strength
with coupling constant $g_M$.  The term $\delta {\cal
  L}_{\rm MSYM}$ in~(\ref{LMSYM}) represents an infinite
number of terms constructed out of $A_i({\bf x})$. We call
this theory {\em Magnetostatic\/} SYM (MSYM) because it is
made out of magnetostatic fields only. The free energy can
therefore be written as
\begin{equation}
  {\cal F} = T \left( f_E 
      + f_M 
      - {\log{{\cal Z}_{\rm MSYM}}\over V} \right)\,,
  \label{FEM}
\end{equation}
where
\begin{equation}
  {\cal Z}_{\rm MSYM} = 
  \int {\cal D}A_i({\bf x})\;
  e^{- \int d^3 x\; {\cal L}_{\rm MSYM}} \, .
\label{ZMSYM}
\end{equation}

We may think of using~(\ref{FEM}) to compute the free energy
density of SYM. We need to determine $f_E$, the effective
parameters of ${\cal L}_{\rm ESYM}$, $f_M$, and the
effective parameters of ${\cal L}_{\rm MSYM}$. The term
$f_E$ and the effective parameters of ${\cal L}_{\rm ESYM}$
may be determined by computing the same physical quantities
in both the full theory (SYM) and in the effective field
theory described by ${\cal L}_{\rm eff}$ (ESYM) and
comparing the result. At first sight this procedure may seem
useless if we consider that to compute physical quantities
in the full theory we have to remove the IR divergences. It
is like having to face the very same problem we wanted to
solve at the beginning. However, to compare results, it is
actually enough to compute them in a region where we know
that both theories describe the same physics.

\bigskip

As mentioned above, self-energy insertions of order $gT$ in
the full theory give rise to IR divergences. If we introduce
an IR momentum cutoff $\Lambda_E$, we will obtain physical
quantities as expansions in powers of $gT/\Lambda_E$. The
expansion diverges when $\Lambda_E$ goes to zero.  However,
if we demand $\Lambda_E\gg gT$, the expansion makes sense.
Of course, the result depends on $\Lambda_E$ and one cannot
get rid of it unless an infinite number of diagrams are
summed up. In the effective theory, the thermal masses
($m_E$ and $m_S$) are also of order $gT$ (we will explicitly
see it below).  Therefore, an IR momentum cutoff
$\Lambda_E\gg gT$ is equivalent to treating the mass terms
in ${\cal L}_{\rm eff}$ as interaction terms. By computing
physical quantities in this way and comparing the results,
we are able to determine the parameters of the effective
theory (ESYM) as functions of $T$, $g$, and $\Lambda_E\gg
gT$.

Once we have determined the parameters of ESYM by matching,
we can use ESYM as an ordinary field theory.  In general,
when computing diagrams using ESYM, we will encounter
ultraviolet (UV) divergences. These UV divergences can be
regulated by introducing a UV momentum cutoff $\Lambda$.
Moreover, ESYM is not fully free of IR divergences. In order
to regulate the IR divergences originated by the gauge boson
self-interaction terms in ESYM, we have to introduce an IR
momentum cutoff $\Lambda_M$.  Analogously to what happened
to the full theory with IR cutoff $\Lambda_E$, we can make
sense of the perturbative expansion of ESYM only if
$\Lambda_M\gg g^2 T$~\cite{lebellac,Nieto:1997pi}.  Note
that the UV cutoff of ESYM has to satisfy
$\Lambda\gg\Lambda_M$. Since $\Lambda_E\gg gT$ and
$\Lambda_M\gg g^2 T$, we can think of $\Lambda_E$ as
$\Lambda_E\sim T$ and $\Lambda_M$ as $\Lambda_M\sim gT$.
Therefore, we can take $\Lambda_E$ as the ultraviolet cutoff
of ESYM: $\Lambda=\Lambda_E$. The $\Lambda_E$ dependence of
the parameters is canceled by the $\Lambda_E$ dependence of
the loop integrals in ESYM.

We see from~(\ref{LMSYM}) that MSYM is a pure gauge theory
in 3 dimensions. It is a confining theory that cannot be
studied perturbatively. Its Feynman diagrams only contain
gauge field self-interaction terms and one should expect the
same IR problems that we have already pointed out for the
full theory and ESYM. However, if we introduce an IR
momentum cutoff $\Lambda_M\gg g^2 T$, we can still use a
diagrammatic expansion to determine the effective parameters
of MSYM by matching physical quantities with ESYM. Of
course, although we may be able to determine the effective
parameters of MSYM, $-\log {\cal Z}_{\rm MSYM}/V$
in~(\ref{FEM}) has still to be calculated
non-perturbatively.

\bigskip 

We write ``$ X \approx \cdots$'' to denote that the physical
quantity $X$ has been computed using the IR cutoffs
$\Lambda_E$ or $\Lambda_M$.  An expression of this type for
$X$ is suitable for matching but it should not be confused
with a {\em correct\/} perturbative expansion (which should
have the IR divergences removed).  The perturbative
expansions performed with IR cutoffs are called {\em
  strict\/} perturbation expansions.

So far, we have used momentum cutoffs to learn how to
organize the perturbative expansion of ${\cal N}=4$
supersymmetric Yang-Mills at finite $T$. However, it turns
out to be more convenient to use a regulator that
automatically gets rid of power divergences. Such a
regulator is dimensional regularization. We therefore
consider the full theory in $(3-2\epsilon)+1$ dimensions.
Since, in the present note, we are interested in computing
the free energy density of ${\cal N}=4$ supersymmetric
$SU(N)$ Yang-Mills to order $\lambda^{3/2}$, we limit
ourselves to the effective parameters and corrections that
contribute to such an order.

The strict perturbation expansions in the full theory can be
greatly simplified if we consider that ${\cal N}=4$
supersymmetric Yang-Mills in 4 dimensions (SYM$_4$) can be
obtained by dimensional reduction of ${\cal N}=1$
supersymmetric Yang-Mills in 10 dimensions~\cite{bss-gso}.
Physical quantities in SYM$_4$ can be obtained
perturbatively by using SYM$_{10}$ Feynman rules while loop
integrals are performed in 4 dimensions (SYM$_{10\rightarrow
  4}$)~\cite{Vazquez-Mozo:1999ic}. In this formulation, the
10d $A_\mu\,(\mu=0,1,2,3)$ can be identified, up to scale
redefinitions, with the 4d gauge field $A_\mu$.  The
remaining $A_{I+3}\,(I=1,\ldots,6)$ are identified with the
6 scalars $\phi_I$ of ${\cal N}=4$ supersymmetric Yang-Mills
in 4d.  While leading order contributions to the {\em
  strict} perturbation expansions in the full theory can be
easily computed using either SYM$_4$ or SYM$_{10\rightarrow
  4}$, higher than leading order contributions are much less
arduous in SYM$_{10\rightarrow 4}$.

\bigskip

The gauge coupling constant of ESYM, $g_E$, can be read off
from the Lagrangian of the full theory. By substituting
$A_0({\bf x},\tau)\rightarrow\sqrt{T}A_0({\bf x})$ in the
Lagrangian of SYM$_4$ and comparing $\int_0^\beta\,d\tau
{\cal L}_{\rm SYM}$ with ${\cal L}_{\rm ESYM}$, we find out,
to leading order in $g^2$,
\begin{equation}
  g_E^2 = g^2 T \, .
\end{equation}

In QCD, the gauge invariant electric screening mass is given
by the pole location of the $\mu=\nu=0$ component of the
complete gluon propagator~\cite{screening}.  We will use a
similar definition in SYM. Using SYM$_{10\rightarrow 4}$,
the inverse gauge field propagator has the form
$\delta^{ab}\left[\delta_{\mu\nu} k^2 + \Pi_{\mu\nu}(0,{\bf
    k})\right]$. In order to compute the static masses in
${\cal L}_{\rm ESYM}$, it is convenient to define $\Pi_{\rm
  el}(k^2)$ and $\Pi_{\rm sc}(k^2)$ such that
$\Pi_{00}(0,{\bf k})\equiv\Pi_{\rm el}(k^2)$ and
$\Pi_{I+3,J+3}(0,{\bf k})\equiv\delta_{IJ}\Pi_{\rm
  sc}(k^2)\,(I,J=1,\ldots,6)$, respectively. Calculations
are performed in Feynman gauge.

\begin{figure}\centering
  \epsfxsize=10cm
  \epsffile{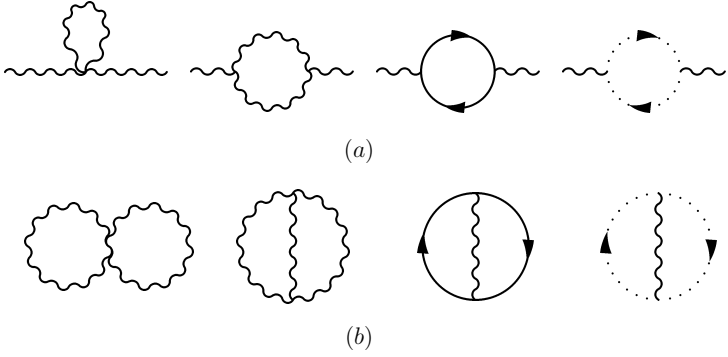}
  \caption{$(a)$ One-loop Feynman diagrams for the gauge field
    self-energy. $(b)$ Two-loop Feynman diagrams for the
    free energy. Wavy lines, solid lines, and dotted lines
    represent the propagators of gauge fields, fermions, and
    ghosts, respectively using SYM$_{10\rightarrow 4}$.}
  \label{fig1}
\end{figure}

The electrostatic mass $m_E$ is determined by computing the
electric screening mass in the full theory and matching it
with the result obtained in ESYM. The electric screening in
SYM is the solution to the equation
\begin{equation}
k^2 \;+\; \Pi_{\rm el}(k^2) \;=\; 0
\qquad \mbox{at $k^2=-m_{\rm el}^2$}\,.
\label{meldef}
\end{equation}
Note that~(\ref{meldef}) gives rise to a double expansion in
powers of $g$ (or equivalently, in number of loops) and in
powers of $m_{\rm el}^2\sim g^2T^2$. In the calculation of
the free energy to order $\lambda^{3/2}$, it is enough to
compute $\Pi_{\rm el}$ at one-loop order. Then, the
expression for the electric screening mass to leading order
in $g^2$ is $m_{\rm el}^2 \;\approx\; \Pi^{(1)}_{\rm
  el}(0)$.  Diagrammatically, $\Pi^{(1)}_{\rm el}(0)$ is
given by the graphs shown in Fig.~\ref{fig1}$(a)$.
These diagrams give rise to sum-integrals that can easily be
evaluated by using standard methods~\cite{lebellac}.  We
find that, in the full theory, the strict perturbation
expansion for $m_{\rm el}$ is
\begin{equation}
  m_{\rm el}^2\approx 2\; C_A g^2 T^2\,.
  \label{melpert}
\end{equation}
In ESYM, the electric screening mass $m_{\rm el}$ gives the
location of the pole in the propagator for the field
$A_0^a({\bf x})$.  Denoting the self-energy function by
$\Pi_E(k^2) \delta^{ab}$, $m_{\rm el}$ is the solution to
\begin{equation}
  k^2 \;+\; m_E^2 \;+\; \Pi_E(k^2) \;=\;0
  \qquad \mbox{at $k^2=-m_{\rm el}^2$}.
  \label{msdefeff}
\end{equation}
Similarly to~(\ref{meldef}), (\ref{msdefeff}) also gives
rise to a double expansion in number of loops and in powers
of $m_{\rm el}^2\sim g^2T^2$. The strict perturbation
expansion of $m_E$ is obtained as an expansion involving
Feynman diagrams evaluated at zero external momentum.  Since
$m_E$, $m_S$, and $g_E$ are treated as perturbation
parameters, there is no energy scale in the dimensionally
regularized integrals and they all vanish.  The solution to
the equation (\ref{msdefeff}) for the screening mass is
therefore trivial: $m_{\rm el}^2 \;\approx\; m_E^2$.
Comparing this result with (\ref{melpert}), we find that,
to leading order in $g^2$
\begin{equation}
  m_E^2 = 2\; C_A g^2 T^2\,.
  \label{ME}
\end{equation}

The calculation of the scalar static mass $m_S$ is similar
to the calculation of $m_E$. We define a scalar screening
mass $m_{\rm sc}$ which satisfies $k^2 + \Pi_{\rm sc}(k^2) =
0$ at $k^2=-m_{\rm sc}^2$ in the full theory and an equation
analogous to~(\ref{msdefeff}) in the effective theory.
After comparing the result in SYM with the result in ESYM,
we find that, to leading order in $g^2$
\begin{equation}
  m_S^2 = C_A g^2 T^2\,.
  \label{MS}
\end{equation}

The effective parameter $f_E$ is evaluated by matching the
calculation of the free energy density in the full theory
and in the effective theory. The leading order contribution
to the free energy density in the full theory is given by a
familiar result from blackbody radiation ${\cal
  F}^{(1)}\approx -\pi^2 T^4/90 \left( \delta_B +
  7\delta_F/8 \right)$, where $\delta_B$ ($\delta_F$) is the
number of bosonic (fermionic) degrees of freedom of the
theory. In the case of ${\cal N}=4$ SYM$_4$,
$\delta_B=\delta_F=8 d_A$, where $d_A$ is the dimension of
the adjoint representation of $SU(N)$: $d_A=N^2-1$.
Alternatively, ${\cal F}^{(1)}$ may be obtained by computing
the one-loop vacuum diagrams of SYM$_4$ using
SYM$_{10\rightarrow 4}$. The next-to-leading order
contribution to the free energy density is given by the
2-loop Feynman diagrams shown in Fig.~\ref{fig1}$(b)$ using
SYM$_{10\rightarrow 4}$.  The evaluation of these diagrams
involves elementary sum-integrals that can easily be
evaluated by standard methods~\cite{lebellac}.  Adding the
2-loop order contribution to ${\cal F}^{(1)}$, we obtain
that the strict perturbation expansion of the free energy
density in the full theory is
\begin{equation}
  {\cal F}\approx -{\pi^2 d_A\over6}T^4\left(
    1 - {3\over 2\pi^2}\; C_A g^2 
  \right) \,.
  \label{fullFE}
\end{equation}

In ESYM the effective parameters $m_E$, $m_S$, and $g_E$ are
treated as perturbation parameters. Then, as there is no scale
in the dimensionally regularized integrals, they all
vanish.  The strict perturbation expansion of the free
energy density is therefore trivial: ${\cal F} \;\approx\; T
f_E$. Comparing this result with (\ref{fullFE}), $f_E$ to
order $g^2$ is
\begin{equation}
  f_E = -{\pi^2 d_A\over6}T^3\left(
    1 - {3\over 2\pi^2}\; C_A g^2
  \right) \,.
  \label{FE}
\end{equation}
As a check we have also computed the effective parameters of
ESYM $m_E$, $m_S$, and $f_E$ using ${\rm SYM}_4$ instead of
${\rm SYM}_{10\rightarrow 4}$ and found the same results
given in Eqs.~(\ref{ME}), (\ref{MS}), and~(\ref{FE}).

To determine $f_M$, we have to compute the strict
perturbation expansion of $-\log{\cal Z}_{\rm ESYM}/V$ and
compare it with that of $f_M-\log{\cal Z}_{\rm MSYM}/V$.
Computing the one-loop vacuum diagrams of ESYM, we find
$-\log{\cal Z}_{\rm ESYM}/V\approx -
d_A/(12\pi)\;(m_E^3+6\;m_S^3)$.  The effective parameters of
${\cal L}_{\rm MSYM}$ are treated as perturbation
parameters. Feynman diagrams give rise to integrals which
have no scale dependence.  Therefore, using dimensional
regularization all the contributions to $-\log{\cal Z}_{\rm
  MSYM}/V$ vanish. We have $f_M - \log{\cal Z}_{\rm
  MSYM}/V\approx f_M$.  Comparing this result with
$-\log{\cal Z}_{\rm ESYM}/V$, we obtain $f_M$:
\begin{equation}
  f_M = - {d_A \over 12\pi}\;(m_E^3+6\;m_S^3)\,.
\end{equation}
The gauge coupling constant $g_M$ can be read off from the
Lagrangian of ESYM ${\cal L}_{\rm ESYM}$. To leading order in
$g_E^2$ and $g^2$, we obtain $g_M^2 = g_E^2 = g^2 T$.

\bigskip 

The free energy of ${\cal N}=4$ supersymmetric $SU(N)$
Yang-Mills is given by~(\ref{FEM}). The term  $-\log{\cal Z}_{\rm
  MSYM}/V$ is non-perturbative. It is easy to find out
at which order non-perturbative corrections become relevant.
The only quantity with dimensions in ${\cal L}_{\rm MSYM}$
is $g_M^2$ (neglecting $\delta{\cal L}_{\rm MSYM}$ whose
contribution is subleading). Since $-\log{{\cal Z}_{\rm
    MSYM}}/V$ has dimension 3 in units of energy, its
lowest order contribution to the free energy is therefore of
order $g_M^6=(g^2 T)^3$.  We conclude that the
non-perturbative corrections generated by $\log{{\cal
    Z}_{\rm MSYM}}/V$ only contribute to the free energy of
SYM at order $\lambda^3$ or higher.

Using~(\ref{FEM}), the free energy of ${\cal N}=4$
supersymmetric $SU(N)$ Yang-Mills in 4 dimensions to order
$\lambda^{3/2}$ is given by ${\cal F} = T\left[ f_E -
  d_A/(12\pi)\;(m_E^3+6\;m_S^3)\right]$.
Substituting~(\ref{ME}), (\ref{MS}), and~(\ref{FE}), we
finally find that, in the large $N$ limit,
\begin{equation}
  f(\lambda) \equiv
   {{\cal F}\over{\cal F}_{\rm ideal}}= 
     1 - {3\over 2\pi^2}\; \lambda +
     {3+\sqrt{2}\over\pi^3}\; \lambda^{3/2}  \,.
\label{F3}
\end{equation}

\bigskip

This result is in agreement with previously reported
calculations. In Ref.~\cite{Fotopoulos:1998es}, the free
energy was evaluated to order $\lambda$.
In~\cite{Vazquez-Mozo:1999ic} and~\cite{Kim:1999sg}, the
term of order $\lambda^{3/2}$ has been computed by
regulating the IR divergences of SYM$_4$ by adding and
subtracting out thermal masses to the Lagrangian.
Vazquez-Mozo~\cite{Vazquez-Mozo:1999ic} used
SYM$_{10\rightarrow 4}$ to compute the first two terms
in~(\ref{F3}).

Since the term of order $\lambda^{3/2}$ in~(\ref{F3}) is
positive, it seems that our result does not favor a
monotonically decreasing function interpolating between the
weak and strong coupling limits. However, such a behavior
may be an effect of the zero convergence radius of the
weak-coupling expansion~(\ref{F3}). It has been
suggested~\cite{Li:1998kd} that there could be a phase
transition in $\lambda$ at large $N$. Even though
perturbative calculations cannot settle the debate on this
matter, it may be possible to find an indication of a phase
transition by using Pade approximants.  An analysis
of~(\ref{F3}) using Pade approximants has been carried out
in~\cite{Kim:1999sg}. There, an indication has been found of
a smooth interpolation between the weak and strong coupling
regimes.

\bigskip

The effective field theory approach provides a systematic
way to compute perturbative contributions to the free energy
of SYM while taking advantage of dimensional reduction
(SYM$_{10\rightarrow 4}$) to determine the effective
parameters of ESYM. This is a major simplification for
computing $f(\lambda)$ to order less than
$\lambda^3$~\cite{nt}. Starting at order $\lambda^3$, there
are non-perturbative contributions as well as perturbative
contributions.  Non-perturbative effects are described by
MSYM, a pure gauge theory in 3 dimensions. As we have
already pointed out, the effective field theory approach
gives rise to effective field theories in 3 dimensions,
without fermions, that are suitable for non-perturbative
studies on the lattice.  Lattice simulations of the
effective field theories could therefore be used to explore
the non-perturbative sector of ${\cal N}=4$ supersymmetric
$SU(N)$ Yang-Mills.

\bigskip

\section*{Acknowledgements}

We are grateful to E.~Braaten and E.~Alvarez for valuable
comments on a draft of this paper. M.T. thanks A.~Wilch
for useful discussions.

\end{document}